\documentclass[preprint]{aastex}
\usepackage{natbib}

\newcommand{\txt}{\textstyle}
\newcommand{\rat}[2]{{\txt #1\over\txt #2}}

\newenvironment{change}{}{}
\newenvironment{changec}{}{}
\newcommand{\ch}{}
\newcommand{\chend}{}
\newcommand{\chb}{\begin{change}}
\newcommand{\chbnd}{\end{change}}
\newcommand{\chc}{\begin{changec}}
\newcommand{\chcend}{\end{changec}}

\slugcomment{Accepted by the Astrophysical Journal}

\shorttitle{Ripples in Solar Magnetic Fields}
\shortauthors{Ulrich and Tran}

\begin{document}


\title{The global solar magnetic field -- identification of travelling, long-lived ripples}


\author{R. K. Ulrich and Tham Tran}
\affil{Department of Physics and Astronomy, University of California,
    Los Angeles, CA 90095}


\begin{abstract}
We have examined the global structure of the solar magnetic field using data from the
FeI spectral line at $\lambda$5250.2\AA\
obtained at the 150-foot tower telescope at the Mt.\ Wilson Observatory (MWO).  For each point on the solar surface, we find the value of the magnetic field in the meridional
plane, $B_m$, by averaging over all available observations using a cosine weighting 
method.  We have revised our cosine weighting method by now taking
into account more fully the highest latitude geometry.  We use the annual variation
in the latitude of the disk center, $b_0$, to deduce the tilt angle of the field 
relative to the local vertical so that we can find the radial component of the field,
$B_r$ from $B_m$.  We find this tilt angle to be small except for
a near-polar zone where a tilt-angle model can reduce the annual variation.  The reduced annual variation in the deduced $B_r$ allows us to study $dB_r/dt$ and associated deviations in $B_r$ from a smoothed $B_r$ with a smoothing width of 2.5 years.  
These functions make evident the presence of small amplitude (3 to 5 gauss) but spatially coherent
ripples with a semi-regular periodicity of 1 to 3 years.  At any given time the
half-wavelength (peak-to-trough) is between 15 and 30 degrees of latitude.  These patterns are 
ubiquitous and in many cases drift from near the equator to the poles over a time
period of roughly 2 years.  The drift rate pattern is not compatible with simple advection.  
\end{abstract}





\section{Introduction}
Magnetic and Doppler measurements of the full solar disk have been obtained 
and recorded digitally at the 150-foot solar tower telescope at the Mt.\ Wilson
Observatory since 1967 (hereinafter the MWO synoptic program, note that earlier observations were recorded by means of various analog systems but these have not been incorporated into the digital record).  Most analyses of these data such as those given by 
\citet{1990ApJ...351..309S,2005ApJ...620L.123U,1988SoPh..117..291U,2002ApJS..139..259U} have relied on a quick-look style database we call the Interactive
Data Reduction (IDR) which includes magnetic and Doppler arrays having spatial dimensions of 34 by 34.  The recent analysis
given by \citet{2010ApJ...725..658U} was based on a new approach which starts from daily average
arrays having spatial dimensions of 256 by 256 wherein each pixel is corrected for differential
rotation prior to its inclusion in the average.  We also include the \citet{2009PhDT.........1T} extrapolation for pixels in the polar regions that are not imaged due to the inclination of the sun's axis of rotation to the
line-of-sight: $b_0$.
In addition we have used a superposed epoch analysis to find a model for the geometry of the solar
surface magnetic field which minimizes the annual variations due to the changing $b_0$. 
Because of these improvements, we are able to study small amplitude variations in the radial component of magnetic field $B_r$. We report here on a new phenomenon that emerges from this cleaned-up data set: a ubiquitous
set of large-scale wave-like features having cycle periods ranging between 0.8 and 2 years. 
\ch Similar features have been found recently by \citet{2012ApJ...749...27V} in the magnetogram data from the US National Solar Observatory.  These features are compared briefly with previously reported periodicities in solar
phenomena, called solar Quasi-Bienniel Oscillations (solar QBO's), but are not studied in enough detail to reveal any physical relationships.\chend

We start with a summary of the magnetic field data and the methods of correction and averaging we have used.
The following section presents the wavelike features and shows that the field patterns are not simply advected
relics of active region.  The final section offers a few thoughts about what these features might
represent.
\section{The Magnetic Field Data}
\subsection{The initial data}
The Babcock magnetograph at the 150-foot tower telescope obtains full-disk magnetograms by scanning the
solar image over the entrance slit and after passage through the spectrograph samples 24 spectral channels 400 times per second in two states of circular polarization.  Averages of the 48 numbers are obtained while the
solar image is moving over the spectrograph entrance slit and stored in a series of records each of which
includes the 48 numbers along with the $x,y$ coordinates of the entrance slit on the solar image as well
as the average time for the observed number.  The observed data are subsequently assembled into a
rectangular grid of pixels of 256 by 256 at a particular time using the differential rotation correction method described by \citet{2006SoPh...235...17U}.  

The raw data numbers obtained from the MWO system are initially stored as 48 polarized intensities along with the red and blue stage Doppler servo positions as functions of 
$x,y,t$ where $x$ and $y$ are the horizontal and vertical positions of the entrance slit on the stage and
$t$ is the average time for the pixel.  We concentrate here on the case of the line at $\lambda$5250.2\AA\ and
use the two spectral samples on opposite sides of that line so that each pixel yields four quantities -- two intensities for each spectral sample.   The four intensities are corrected for blue and red stage servo error and converted according to center-to-limb dependent calibration curves into a magnetic signal and a Doppler signal.  These steps are
described in greater detail in \citet{1983SoPh...87..195H} and \citet{2002ApJS..139..259U}.  Our applications of the data to modeling of the Total Solar Irradiance (TSI) \citep{2010SoPh..261...11U} and meridional circulation \citep{2010ApJ...725..658U} have demonstrated the effectiveness of using daily averages and limiting the days used to those days having three or more observations and we have adopted that approach in the present analysis.
Prior to 1986 there were few days that satisfied this condition and all good observations were included in
the analysis.

The observations are made using one of two entrance aperture, slow-grams with a 12 arc-sec squared aperture and fast-grams with a 20 arc-sec squared aperture.  Comparison of the magnetograms shows that the fast-grams have a systematically smaller field than the slow-grams.  Our program has a mix of slow- and fast-grams and we find that the systematic effect can be removed by dividing the field from the daily average $(B_{\rm los})_{\rm ave}$ by a factor $C_{\rm fs}$ to get an equivalent field that would have been measured as a slow-gram $(B_{\rm los})_{\rm slow}$.  We find that $C_{\rm fs}\approx 1-0.0127\,N_{\rm fast}$ where $N_{\rm fast}$ is the number of fast-grams on that day.  Typically $N_{\rm fast}=N_{\rm total}-2$ where $N_{\rm total}$ is the number of observations of both types during the day.
In addition, we have multiplied the observed magnetic fields by the correction factor by \citet{2009SoPh..255...53U} to bring the field strength to the value appropriate for the base of the chromosphere.  As discussed in the above reference this factor corrects for effects due to the fluxtube nature of the photospheric magnetic field.

The array size we use for this analysis is 256$\times$256 yielding a pixel size of about 8 arc-sec.  This
is smaller than the actual pixel size but due to the dithering effect of adding together multiple observations,
this pixel size is close to an effective resolution element.  We have assembled various results from the analysis into latitude dependent arrays by averaging into bins $2^\circ$ wide with the bins being centered on odd values of latitude (this avoids dividing by the sine of the latitude at the equator and by the cosine of the latitude at the pole).  We will generally use degrees instead of radians to discuss the dependency of $B_r$ on latitude.

\subsection{Calculation of the radial field from the line-of-sight field}
The deduction of the magnetic field geometry has been the topic of a number studies such as those by 
\citet{1978SoPh...58..225S}, \citet{1992ApJ...392..310W} and \citet{2009ApJ...699..871P}.  We describe
here a new method that depends on the long duration of the digital record from the MWO synoptic program.
Following an approach similar to that described by \citet{1994SoPh..153..131S} we use solar rotation to resolve the line-of-sight magnetic field $B_\ell$ into components in the meridional plane $B_{\rm m}$ (the plane through the observed point and the solar axis of rotation) and in the zonal plane $B_{\rm z}$ (the plane parallel to the solar equator).  Our method is described most completely by \citet{2006SoPh...235...17U} and results from our method have been presented at various times starting with \citet{1993ASPC...40...25U}.  The essential ingredient in our approach and the Shrauner-Scherrer method is to weight the observed line-of-sight field $B_\ell$ by the sine and cosine of the central meridian angle $\delta L$ and sum over all available observations.  The meridional and zonal components then can be recovered from a matrix inversion.  We have found that the inclusion of sums of cross terms involving the products $\cos(\delta L)\sin(\delta L)$ magnifies errors in cases where the data is sparse.  For uniformly distributed data, these terms should vanish but in our actual data sets, they do not.  We find the final result is improved by setting
such cross terms equal to zero.

In this paper we are interested in the magnetic field at high latitudes and a term neglected previously needs to be included.  Equations relating magnetic field components in a rectangular grid for an observed image to
the components in a heliographic coordinate system have been given by \citet{1987SoPh..107..239H}.  Because
our observations are made after we have determined the east-west direction at the focal plane and oriented
the scan lines so they \ch are \chend perpendicular to the sun's axis of rotation, our images always have the $p$ angle effectively set to zero.  Also we only measure the line-of-sight field component $B_\ell$.  The Hagyard equations can then be written:
\begin{eqnarray}
B_\ell&=& \left[(B_r\cos b+B_\theta\sin b)\cos\delta L - B_\phi\sin \delta L\right]\cos b_0
+(B_r\sin b-B_\theta\cos b)\sin b_0
\label{eqnone}
\end{eqnarray}
where $b$ is the latitude, $b_0$ is the latitude at disk center, $B_r$, $B_\theta$ and $B_\phi$ are respectively the radial component of the magnetic field, the component of the field in the colatitude direction and the
component of the field in the $\phi$ direction for a right-handed spherical coordinate system.  We define a
tilt angle \ch $\zeta=\arctan(B_\theta/B_r)$ \chend giving the angle of the field line with respect to the local vertical measured positive from the north polar axis toward
the south polar axis.  We describe below how a model for $\zeta$ as a function of $b$ can be found by 
minimizing the variance of $B_r$.  Using $\zeta$ equation(\ref{eqnone}) becomes:
\begin{eqnarray}
B_\ell\cos\zeta&=&\left[B_r\cos(b-\zeta)\cos\delta L-B_\phi\cos\zeta\sin\delta L\right]\cos b_0+B_r\sin(b-\zeta)\sin b_0\ .
\end{eqnarray}
The next step is to multiply by $\cos\delta L$ and sum over all available observations.  In accordance with
our modification of the Shrauner-Scherrer method, we set $\sum \cos\delta L\sin \delta L=0$ and $\sum \sin\delta L=0$ and assume that $B_r$ is constant over the set of observations.  This gives:
\begin{eqnarray}
B_r&=&\rat {\cos\zeta\sum B_\ell\cos\delta L}{\cos(b-\zeta)\cos b_0\sum (\cos\delta L)^2 + \sin(b-\zeta)\sin b_0
\sum\cos \delta L}\ .
\label{eqnthree}
\end{eqnarray}

\subsection{Filling in unseen portions of the polar regions}
\begin{figure}
\begin{center}
\parbox{6.9in}{\begin{center}
\resizebox{5.8in}{!}{\includegraphics{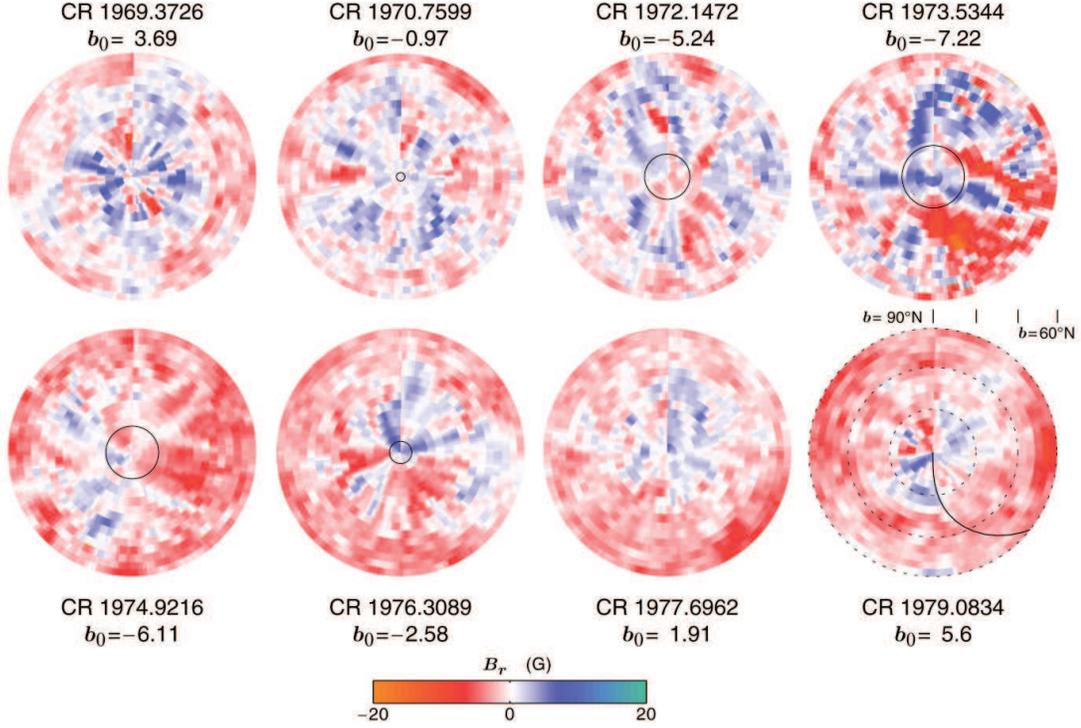}}
\caption{This figure shows a sequence of projections of maps of the radial component of the sun's magnetic field.  Each circle shows $B_r$ as a function of position for a Carrington time value as indicated
below the circle.  The Carrington rotation number and corresponding Carrington longitude apply to the line
extending vertically down from the pole point to the lower edge of the circle.  For other angles the heliographic longitude can be computed from angle difference from this vertical line.  This longitude is not the Carrington longitude for any other angle on the circle.  All points on the figure have been corrected to correspond to the labeled Carrington time according to the methods given by \citet{2006SoPh...235...17U}.  The latitude as projected is given above the lower right circle and is shown on this figure
by the dashed circles.  Each circle
is confined to latitudes pole-ward of $60^\circ$.  The circular solid lines show the boundary of the zone where the field value must be determined from our polar fill technique.  Each circle is offset from the previous by an interval that corresponds to the rotation period at the pole.  The curve on the lower right circle gives the pattern that would be produced on that circle if a vertical straight line descending from the pole point had been drawn on the preceding circle.  This curve can be used to estimate which features might have been preserved from one polar rotation to the next.}
\label{figureonea}\end{center}
}
\end{center}
\end{figure}

A new step in our reduction is the filling of the unseen parts of the polar regions with a method based on the 
ideas developed by \citet{2009PhDT.........1T}.  \ch The problem of estimating the magnetic
field for those portions of the disk that are beyond the limb has been addressed in a
variety of ways and having a good approach is important for the polar regions whose
dipole-like field influences a major portion of interplanetary space.  A variety of
methods have been developed and tested with an excellent recent discussion by \citet{2011SoPh..270....9S} which
includes a summary of seven methods and introduces a new extension based on a combination
of the best features of several of those methods.  The method we present here differs
from all of those by transforming the image to a $b_0=0$ configuration prior to the step
of deriving the fill values. The study by \citet{2009PhDT.........1T} has shown that
this approach avoids generating field structures aligned along lines of constant longitude
which are seen in some polar fills based on polynomial interpolation.  The Tran method
as we have implemented it is stated below; \chb and we also illustrate the performance of the method using samples of individual maps as viewed from the pole.  \chbnd  A detailed comparison between this method
and other methods is, nonetheless, beyond the scope of this paper. 

\chb The Tran method is particularly well suited for application to the MWO data because 1) the error of measurement from the MWO system is substantially less than 1 G \citep{2009SoPh..255...53U}, 2) the scanning aperture method of measurement prevents scattered light within the spectral analyzer from influencing the field strength near the limb in contrast to filter-based system where scattering from bright parts of the image can reach the limb, and 3) the signal is sampled at 400 Hz so that the seeing fluctuations are cancelled in the algorithm that provides the magnetic signal.  An empirical estimate of the noise is difficult and our best resource is a set of images taken periodically with the circular polarization analyzer turned off.  The distribution of apparent magnetic signal from these pseudo-magnetograms gives a gaussian width of 0.5 G.  Each daily average is composed of an average of 11 individual magnetograms while near the limb, the fluxtube physics correction factor discussed in \citet{2009SoPh..255...53U} multiplies the raw field strength by about 1.8 and the conversion from line-of-sight to radial multiplies the field by another factor of 3.  The combination of these factors gives an error of about 1 G in $B_r$ for each pixel on the image near the limb.\chbnd\  

To calculate fill values we take the observed values of $B_r$ derived from the
geometry described above and rotate the image so that the equator is at the center of the disk.  The observed
pixel values then occupy positions on this image different from where they were observed.  In particular pixels
near the pole do not have observed values and these are the pixels for which we seek appropriate estimated values.  For all such pixels, we find on this rotated plane image the six nearest pixels with observed values.
We then calculate the projected distance on this rotated image between each observed pixel and the pixel for
which we wish to find an estimate.  We take for this estimate the weighted average of the observed pixels with the weighting function being the inverse distance between the target pixel and each observed pixel on the projected and rotated image.  This method allows longitudinal structures to influence the nearby polar pixels in a natural and bounded manner.
In the $b_0=0$ geometry, the nearest pixels as defined this way include a range in longitude which increases as the pixel is further from the edge of the observed image so that there is
less longitudinal concentration of the nearby structures.\chend

\chb In order to illustrate the performance of the pole filling approach, we have selected times near polar field reversal to form eight snapshot maps of the radial component of the solar field.  The coordinates on
these maps are heliographic longitude and latitude.  Each pixel is corrected for differential rotation and averaged according to equation~(\ref{eqnthree}).  The map has been reprojected so that it appears as if being
viewed from above the sun's north pole.  Rotation is in a counter-clockwise direction and the time of all points is corrected to that on the central meridian (a line vertically downward from the circle center).  Each circle shown only includes latitudes northward of 60$^\circ$.  The format of the figures have been chosen to agree with that used by \citet{2011SoPh..270....9S} although the times chosen do not correspond to those used by those authors.  The time of observation is given in Carrington time units \citep{2006SoPh...235...17U} and is listed above or below each circle.  The corresponding times in fractional years are 2000.849, 2000.952, 2001.056, 2001.160, 2001.264, 2001.367, 2001.471 and 2001.574.

We see from the sequence in figure~(\ref{figureonea}) that the structures in the polar region during this
stage of the cycle are quite variable from one rotation to the next.  Our method of calculating $B_r$ as given in equation~(\ref{eqnthree}) can artificially enhance features in some longitude ranges as a result
of irregularly space observations.  Some of the strong patterns in the upper left and upper right circles may be due to this effect.  However, the extension into the unseen areas in the upper right circle is not enhanced beyond the features in the seen areas and no fill has been applied to the upper left figure.  In general the fill pattern is a logical extension of the seen patterns with no evident enhancement of noise patterns.

\chbnd\

\subsection{Determination of the tilt angle $\zeta$}

\begin{figure}
\begin{center}
\parbox{6.5in}{
\resizebox{6.5in}{!}{\includegraphics{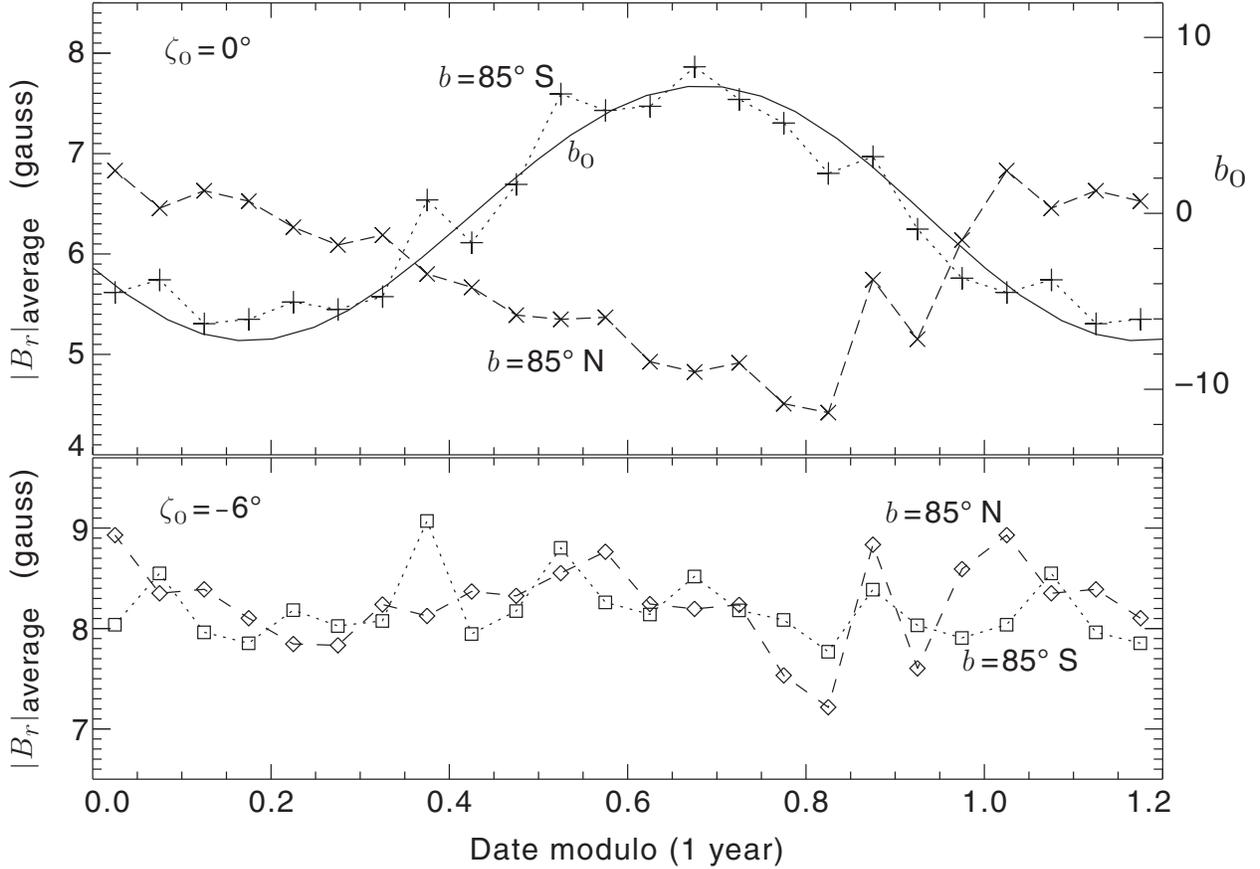}}
\caption{This figure shows a superposed epoch analysis of $B_r$ as derived from equation~(\ref{eqnthree}) and averaged over the full data set.  The superposed epoch average is obtained by adopting the independent 
variable to be the time modulo 1 year.  The portion of the figure between 1.0 and 1.2 year is a copy of the function between 0.0 and 0.2 and is included to facilitate examination of the
trends.  The top panel has the tilt angle $\zeta$ taken to be zero so that the assumed field orientation
is exactly vertical.  A selected pair of latitudes at 85$^\circ$N and 85$^\circ$S are shown along with
a plot of the apparent polar tilt angle $b_0$ as the solid line.  The scale of $b_0$ and location of the line has been adjusted to aid in the comparison to the case of $b=85^\circ$S.  The lower panel shows the two latitude
results after a model for $\zeta$ as described in the text is applied.
}
\label{figone}
}
\end{center}
\end{figure}

For an analysis in which $\zeta$ is assumed zero (the field lines are exactly vertical everywhere), we find
there to be a strong annual variation in the derived $B_r$, especially in the polar regions where there are
long periods during which one magnetic polarity or the other dominates.   The influence of $b_0$ comes from the
denominator of equation~(\ref{eqnthree}) which for $\delta L$ near zero is just $\cos(b-\zeta-b_0)$.  As $b-\zeta-b_0$ approaches $\pm90^\circ$, the line-of-sight magnetic field is multiplied by a factor that can be quite large.  We 
isolate this effect by considering an annual superposed epoch analysis whereby all the derived values of $|B_r|$ are treated as functions of the time of the year and averaged over the full data set.  To illustrate the need
for inclusion of the tilt angle $\zeta$ we show in figure~(\ref{figone}) a sample superposed epoch result for
latitudes $b=85^\circ$N/S.  The top panel shows the result for vertical fields with $\zeta=0$ everywhere.  At 
the latitude chosen, $\cos(b-b_0)\approx\sin(b_0)$.  For comparison we show $b_0$ in this top panel with the
scale adjusted in amplitude and offset to coincide with the superposed epoch result for ${B_r}_{\rm average}$ at the south latitude.  

\begin{figure}
\begin{center}
\parbox{6.5in}{
\resizebox{6.5in}{!}{\includegraphics{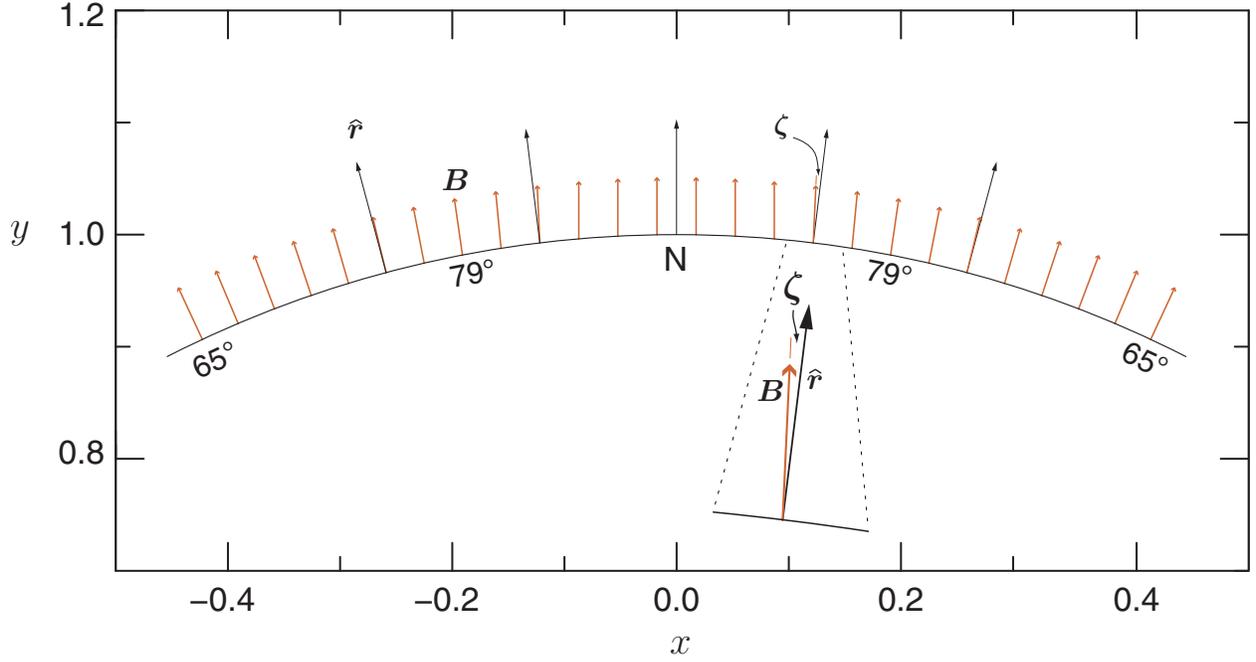}}
\caption{This figure shows the geometry of the adopted field line tilt as defined by the tilt angle $\zeta$.  Each of the arrows is in the direction of the field lines but for purposes of illustration are shown with
a length of 5\%\ of the solar radius.  For reference, the longer arrows show the direction of the local vertical. Each of the shorter arrows is at the center of one of the bins used to describe
the magnetic field pattern.  The $\zeta$ character with the curved arrow points to a place where the tilt angle 
has the largest value.  The $x$ and $y$ coordinates are in units of the apparent solar radius and are oriented in the same way as the image is observed.  The expanded inset shows the definition of $\zeta$.}
\label{figtwo}
}
\end{center}
\end{figure}

\begin{figure}
\begin{center}
\parbox{6.5in}{
\resizebox{6.5in}{!}{\includegraphics{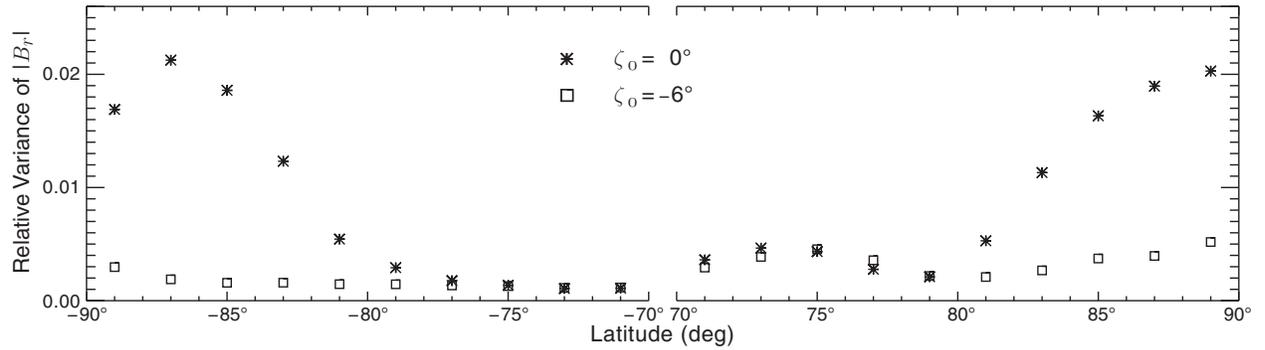}}
\caption{This figure shows the variance of the superposed epoch averages of $|B_r|$ (see figure[\ref{figone}]) as a function of latitude in the polar regions for the case where the field is assumed vertical everywhere and for the case where the field obeys the model given by equation~(\ref{eqnfour}).}
\label{figthree}
}
\end{center}
\end{figure}

We can explain the nature of the pattern in the top panel of figure~(\ref{figone}) by concentrating on the plot for the southern hemisphere where there is greater regularity.  The sense of the variation is such that when $b_0$ is the most negative and $85^\circ$S is most visible, the deduced field is relatively weak.  When the latitude is least visible and is filled in by regions slightly to the north, the result is stronger.  On the adjacent latitude of $77^\circ$S where filling by extrapolation is not needed, the result is a relatively constant $6.5$G without the $2.5$G annual variation seen at $85^\circ$S.  Interpreted literally this behavior implies that there is a decrease in the field strength at the pole and that the field strength gradient depends on the time of the year.  We believe that the correct interpretation is that the field is not radially oriented near the pole and that the correct from $B_\ell$ to $B_r$ requires a larger multiplicative factor -- hence the field is tilted slightly away from the line-of-sight and is more nearly perpendicular to the line-of-sight.  We impose an assumption that the field lines do not converge toward each other; but, if some model were to suggest that this assumption is invalid, it would not impact the effectiveness of our analysis.  This assumption imposes the condition that $\zeta>b-90^\circ$ for $b>0$ and $\zeta<b+90^\circ$ for $b<0$.  As a convenience in developing our model, we have used the quantity \ch$\xi=90^\circ-|b|$ \chend to represent the proximity to the poles.  The model for $\zeta$ we have adopted is:
\begin{eqnarray}
\zeta&=& {\sf sign}(b)\;{\sf max}\left[\zeta_0\exp\left(-(\xi/\delta\xi)^2\right),|b|-90^\circ\right]
\label{eqnfour}
\end{eqnarray}
where ${\sf sign}=1\;\hbox{for}\;b>0\;\;\hbox{and}\;\;{\sf sign}=-1\;\hbox{for}\;b<0$ and where ${\sf max}[a,b]$ is the maximum of $a$ and $b$.  We have adopted $\zeta_0=-6^\circ$ and \ch$\delta\xi=12.5^\circ$\chend.  Figure~(\ref{figtwo}) illustrates the geometry of the field lines near the sun's north pole.  The improvement in the variance of $|B_r|$ is illustrated in figure~(\ref{figthree}).


\subsection{Relationship with the fluxtube correction factor}
\label{fluxtube}
As indicated above, the analysis presented here has been applied to the line-of-sight magnetic fields that
have been corrected for the effects of photospheric fluxtube physics according to the recommendation given by \citet{2009SoPh..255...53U}.  In keeping with the traditional approach with MWO data, we initially worked with the uncorrected fields.  With that approach, we found $\zeta$ to be near zero close to the poles and to have a maximum of about 13$^\circ$ for latitudes near 65$^\circ$N/S.  We were surprised to find that after the application of the fluxtube correction factor, the field became radial almost everywhere with the exception being the very high latitude regions presented above.  We checked to determine if the effect could be due to an error in the treatment of the near-limb pixels around the whole circumference of the solar image.  This last assumption failed to remove the near-pole effects because the feature we correct is actually a contrast between properties of the solar surface at latitudes greater than 80$^\circ$ compared to those between 65$^\circ$ and 80$^\circ$ and those latitudes are altered equally in the case of a full-circumference correction.  \ch The character of the center-to-limb dependence of
the field strength is discussed further in the subsection below.\chend

\begin{figure}
\begin{center}
\parbox{6.5in}{
\resizebox{6.5in}{!}{\includegraphics{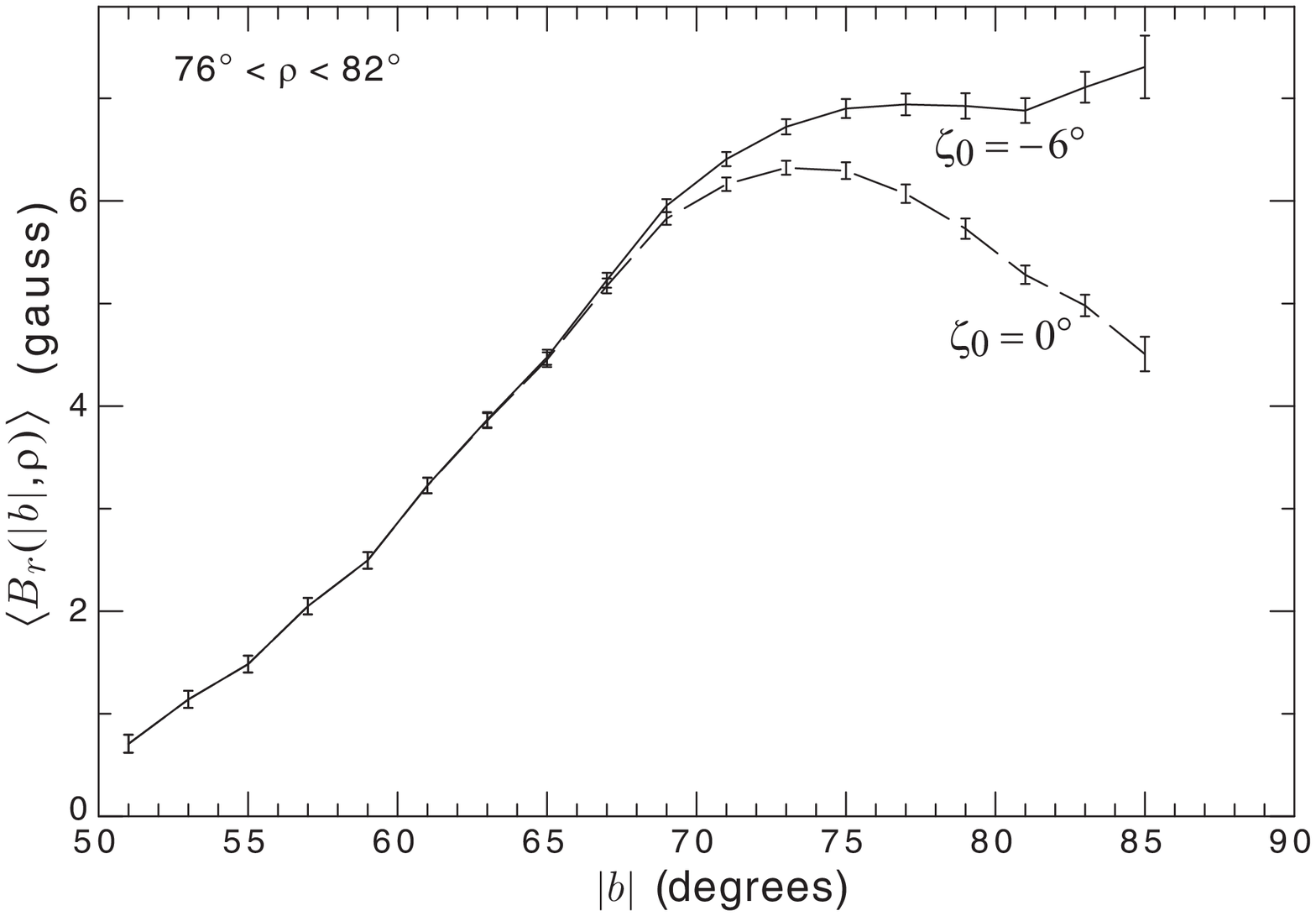}}
\caption{This figure shows the latitude dependence of the latitude restricted center-to-limb function as defined in the text.  The points averaged together are all observed and not
a result of the fill procedure.  
{
The error bars are the errors of the mean derived from the variance of the data points included for each point.}  The two lines show averages from deduced values of $B_r$ based on two models of the field geometry: the first with a strictly radial assumption (the dashed line) and the second with the poleward tilt implied by the model discussed in the text (the solid line).}
\label{figthreea}
}
\end{center}
\end{figure}

\ch
\subsection{Influence of the center-to-limb angle}
Although the influence of the $b_0$ angle on the near-polar field is most strongly
seen in the polar fill zone, in fact the dependence of the fully seen pixels on
both the latitude $b$ and the center-to-limb angle $\rho$ illustrates the need for
the $\zeta_0=-6^\circ$ model we have adopted.  Due to the long temporal record available to us, we can study the systematic dependence of the field on both these variables by collecting all observed points into bins of $b$ and $\rho$.  We have taken the bin width
to be 2 degrees in extent centered on odd values in degrees.  

The simplest approach is to average the absolute values of the $B_r$ but this approach produces an artificial increase in $|B_r|_{\rm average}$ due to the noisiness of the points nearest to the limb where the noise can be comparable to the value of $B_r$.  We can overcome this effect by using the fact that in the near-polar zone of 70$^\circ$ to
90$^\circ$ and in the high-latitude zone of 50$^\circ$ to 70$^\circ$, the field
is mostly positive or negative between the times of polar field reversal.  Thus we can
average the signed value of $B_r$ and obtain nearly the same result as averaging the
absolute value as long as we reverse the sign whenever the polar region has a negative
predominant polarity.  Then when the noise is comparable to the value, we will not
get an increase in the average because the positive and negative offsets due to noise
will average to zero.  The dates we have adopted are: prior to 1981.0 the sign in
the south is reversed, between 1980.4 and 1990.2 the sign in the north is reversed, between 1991.5 and 2000.7 the sign in the south is reversed and after 2000.7 the sign in the north is reversed.  These times are appropriate for 70$^\circ$.  Because the fields are small near these reversal times, the shifts in time of reversal for other nearby latitudes do not impact the averages.  In addition, after average is computed for both north and south latitudes, we average together results for equal values of $|b|$ and call the result the Latitude Restricted Center-to-Limb Function (LRCtLF): $\left<B_r(|b|,\rho)\right>$.  The restriction in latitude also restricts the range of possible values for $\rho$ since the
closest any latitude gets to the disk center at any time of year has $\rho>|b|-b_0$. 

Because the LRCtLF isolates observations having the same values of $\rho$, the resulting
latitude dependence can only come from the combination of magnetic field strength and magnetic field geometry.  While we cannot separate these two factors, we can compare the implied functions based on the $\zeta$ model.  In particular we examine the cases $\zeta_0=0^\circ$ and $\zeta_0=-6^\circ$.  Figure~(\ref{figthreea}) shows that the strictly radial field 
assumption results in a noticeable reduction in field strength at the highest latitudes
whereas the model with the poleward tilt shows a pattern of field strength which increases
smoothly toward the pole.  

For the highest latitudes, the LRCtLF is difficult to use to study the $\rho$ dependence
because $\rho$ can only have values between 90$^\circ$ and $|b|-7.25^\circ$ so the functional pattern of the dependence
of $B_r$ on $\rho$ cannot be studied.  In addition we have too few points for $\rho>82^\circ$ to provide a statistically valid result.  The variance test of the superposed epoch values of $|B_r|$ includes the result of the fill technique and brings in information from lower latitudes.  Over the restricted range of $\rho$ available for the observed points as opposed to the result including the fill points, we find that from lowest $\rho$ to highest $\rho$, the LRCtLF decreases by 11\%\ with $\zeta_0=-6^\circ$ and 23\%\ with $\zeta_0=0^\circ$.  This smaller change in $B_r$ with $\rho$ is responsible for the lower variance in the superposed epoch function.  We have tried modifications to the $\rho$ dependence of the field strength
similar to those applied above in section~\ref{fluxtube} and found that changes causing
$\zeta_0$ to become near zero also result in the need for positive $\zeta$ in latitudes
below the near-polar zone.  We cannot rule out such models but feel the present model
is simplest in that it requires only one adjustment - the near-polar $\zeta$ model instead
of two - an adjustment to the fluxtube physics correction function and a lower latitude
model for $\zeta$.  We note, however, that the profile for the FeI line at $\lambda5250.2$\AA\ has not been measured with our system with $\rho$ above 70$^\circ$ so that some
adjustment to our result could be necessary based on improved observations of this line 
profile.  
\chend

\section{The Moving Magnetic Patterns}

\begin{figure}
\begin{center}
\parbox{6.5in}{
\resizebox{6.5in}{!}{\includegraphics{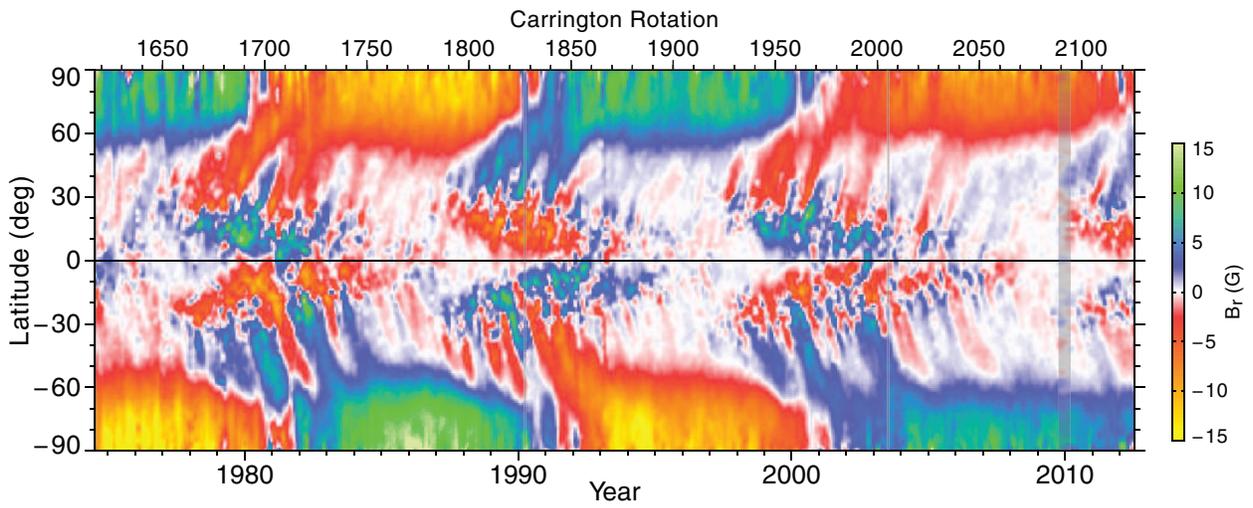}}
\caption{This figure shows the strength of the radial component of the magnetic field as a function of
latitude and time.  Time is given in years along the lower axis and in Carrington rotation number along the top axis.  The color-bar to the right shows the correspondence between the colors and the strength of the radial component of the magnetic field with the units of the field (flux density) being in Gauss.  Three gaps in the data due to weather conditions and equipment failures have been filled by means of a cubic spline interpolation at locations indicated by the grey rectangles near 1990.2, 2003.5 and 2010.}
\label{figurefour}
}
\end{center}
\end{figure}

Long-term trends in the solar magnetic field are usually shown in 2-d plots giving the field strength as a 
function of time and latitude.  The MWO project has published a number of these as derived from the IDR 
records whose basic spatial resolution starts with a grid 34$\times$34 in sin(central-meridian angle) and sin(latitude).  An example is found in \citet{2005ApJ...620L.123U}.  As described above, our current reduction differs from those previously as a result of starting with a spatial resolution of 256$\times$256 as derived from daily averages of all
images available.  The prior approach yields arrays in latitude that are chosen so that the spacing is
roughly uniform in $\sin(b)$.  That produces a coarse result at high latitudes where the bins at indices
1 and 34 include areas between the poles and 76$^\circ$N/S.  In addition, the prior plots show the raw, line-of-sight magnetic field instead of the radial field after the correction for flux-tube effects.  Two consequences of this change are 1) the field in the polar regions is now significantly stronger than in earlier plots and 2) the annual effects seen before are now largely absent.  Furthermore, our adoption of the Tran polar extrapolation method provides an improved level of detail in the time dependence of the polar fields.
Our new map of the radial component of the magnetic field flux density $B_r$ as a function of latitude and time is given in figure~(\ref{figurefour}).

\ch
The new map is largely free of the annual variations produced from the variation of $b_0$, however there are a few features worth comment.  First, there are sequences like those near the north pole between 1992 and 1995 where a four successive local minima are separated by one year.  However, this sequence and others like it do not persist and do not influence the overall superposed epoch values.  Second, 1 to 2 years prior to each cycle's final polar field reversal there is a preceeding
mini-reversal that does not remain.  These are consistent in time relative to the reversal even for
cases where the north and south reversals are themselves out of phase.  The color mapping we use makes these features more evident because we make the color gradient greater near zero field in order to better display weaker fields in the mid-latitudes.  In the polar regions when the field is reversing, this color mapping combined with the magnetic field ripples discussed below produces an effect which appears to be a mini-reversal preceeding the main reversal by one to two years. \chend

\begin{figure}
\begin{center}
\parbox{5.0in}{
\resizebox{5.0in}{!}{\includegraphics{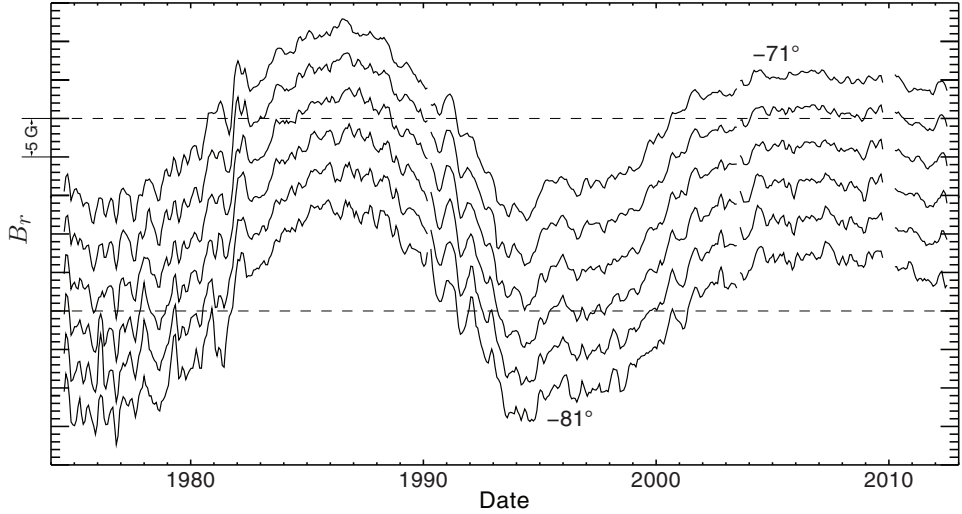}}
\caption{The time dependence of the radial component of the magnetic field $B_r$ for six selected latitudes near the sun's south pole.   The dashed lines across the figure correspond to zero field for the top and bottom latitudes.  Each latitude line is offset by 5 G from the adjacent line.
}
\label{figurefive}
}
\end{center}
\end{figure} 
\begin{figure}
\begin{center}
\parbox{6.5in}{
\resizebox{6.5in}{!}{\includegraphics{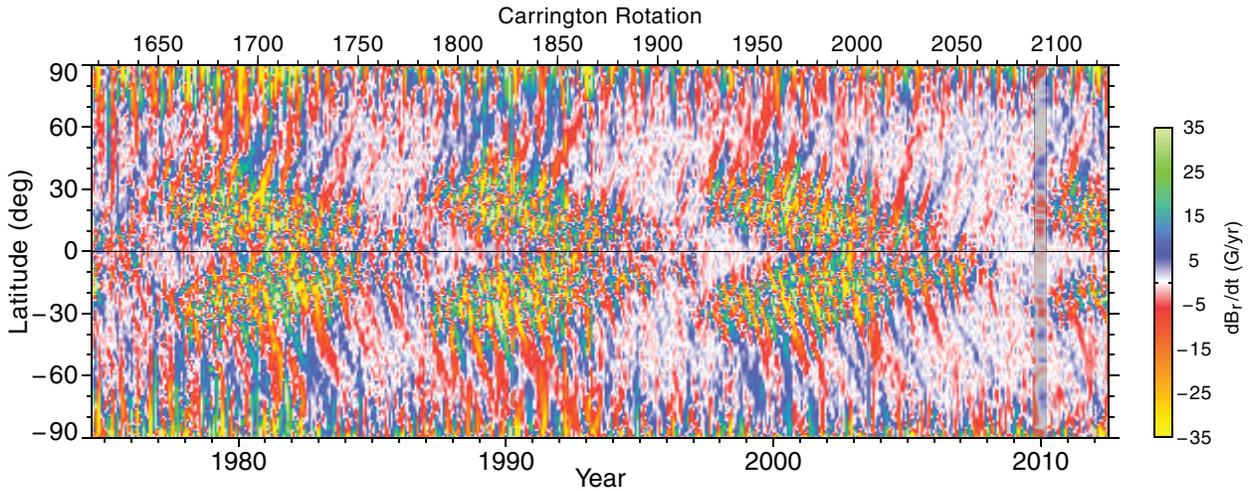}}
\caption{This figure gives a map of the time derivative of the radial magnetic field $dB_r/dt$ in units of
G/yr.  The color-bar on the right gives the coding of the mapped colors.  The magnetic field for this
mapping has only been averaged over a single Carrington rotation.  Gaps in the record have been filled using spline interpolation with the filled portions being indicated as in figure~(\ref{figurefour}).}
\label{figuresixa}
}
\end{center}
\end{figure}
\begin{figure}
\begin{center}
\parbox{6.5in}{
\resizebox{6.5in}{!}{\includegraphics{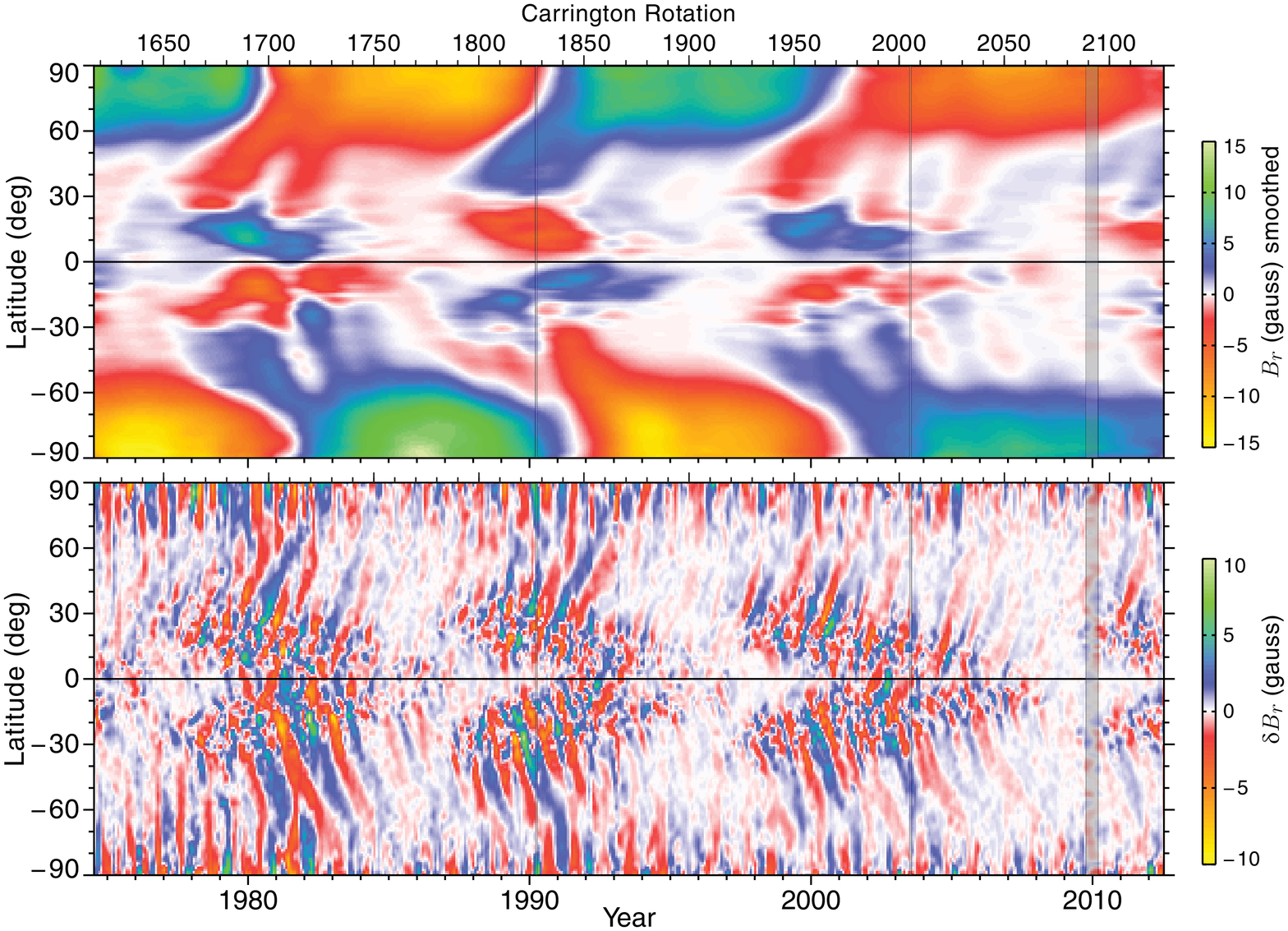}}
\caption{This figure shows $B_r$ smoothed in the top panel and the difference $\delta B_r$ between $B_r$ and the smoothed $B_r$ in the lower panel.  The smoothing function is a truncated gaussian with a width of 2.5 years.
  Gaps filled
by spline interpolation are indicated on both panels as in figure~(\ref{figurefour}).
}
\label{figuresixb}
}
\end{center}
\end{figure}

An important property of the $B_r$ map is the regular occurrence of magnetic plumes that start near the active regions or near the equator and migrate poleward.  These are easily recognized in maps of this sort and play a critical role in the reversal of the polar magnetic field near sunspot maximum.  What has been less evident is their ubiquitous presence at all phases of the solar cycle.  In areas of weaker field, the regular, shorter-term ripples would be lost without the consistent quality of figure~(\ref{figurefour}).  These ripples can be seen more clearly by examining a subset of line plots of $B_r$ vs.\ $t$ as are shown in figure~(\ref{figurefive}).  Although the $2^\circ$ bin size oversamples the MWO images, the consistent appearance and time drift associated with the features indicates that they are not simply random noise.

The ripples can be displayed effectively by examining $dB_r/dt$ since this operation enhances the short-term changes relative to the 11-year solar cycle pattern.  Initially we were searching for times of rapid field change in order to find patterns in the Doppler velocities.  We believed that the small variations in the field were in fact noise so that we reduced these by smoothing.   
\ch Subsequently we determined that the features are coherent over long time periods so that we now show the results of taking the derivative on the non-smoothed instead of the smoothed fields.  We have used the non-smoothed fields shown in figure~(\ref{figurefour}) to take numerical derivatives of the $B_r$ trend lines.  The resulting time derivative map is shown
in figure~(\ref{figuresixa}).   The rough regularity of the pattern in figure (\ref{figuresixa}) was unexpected and represents a behavior that deserves attention.  The continuity of the features from low to high latitudes is particularly striking and is found for even relatively weak ripples.

The overall pattern of the ripples brought out by the time derivative is related to the structures
found by \citet{2012ApJ...749...27V} who employed an iterative smoothing/differencing method called Empirical Mode Decomposition to isolate what are called Intrinsic Mode Functions (IMF's).  The process includes spline fitting to maxima and minima of the starting function and a set of tests to 
determine when an Intrinsic Mode Function has been accurately found.  The resulting functions are 
collected together in groups to show time and space dependence of the magnetic field.  The choice
of IMF's to group governs the range of frequencies retained.  The authors show two groups, one with frequencies near one year and the other with frequencies between 1.5 and 4 years.  Although both group results resemble our maps based on $dB_r/dt$, there are differences that are significant such as the patterns that drift toward the equator on their maps -- a drift which is not seen on our maps.  We considered the use of spline fitting functions to remove the long-time
background variations from our $B_r$ maps in order to display the ripples as magnetic field strength variations instead
of using the time derivative but we found that the spline fit approach can introduce artifacts due to the stiffness of the fit function and prefer the method described in the following paragraph.

We have adopted the difference $\delta B_r$ between the raw $B_r$ (figure~\ref{figurefour}) and a smoothed $B_r$ (figure~\ref{figuresixb}, top panel) as a simple and effective way of representing the ripples.  Figure \ref{figuresixb} shows the result when we use a smoothing function with a truncated gaussian of width of 2.5 years.  This choice of smoothing effectively divides the function into a primary wave, dipole-like component and a ripple component.  At any fixed time, the $\delta B_r$ function represents a perturbation on top of the primary component.  The near-sun heliosphere is generally considered to be current free and describable by potential field solutions \citep{1969SoPh....9..131A,1992ApJ...392..310W}.  Consequently, the response of the field above the photosphere is linear to variations in the observed field.  
\chb  We can then consider the heliospheric field in this region to be the sum of three parts: the smoothed primary wave, a superposed response from the ripples and finally the longitude-dependent component that includes the effects of active regions and sunspots.  The three vector components of the field include contributions from each of these components.  If the ripples acted in isolation this response would take the form of loops whose foot points are separated by 15$^\circ$ to 30$^\circ$ of latitude and which would migrate mostly toward the poles.  Compared to the smoothed wave, the amplitude of the loops is about 30\%\ of the primary dipole-like field, at least during periods of high activity.  However, the actual trace of the field lines is dominated by the more structured longitude-dependent component which is the strongest of the three.  The effect of the ripples is to tilt the field lines so that they go in the direction of the loops that would be present if the longitude-dependent structures were missing.  Detailed models are needed to determine if this effect is present.
\chbnd

\chend

\begin{figure}
\begin{center}
\parbox{6.0in}{
\resizebox{6.0in}{!}{\includegraphics{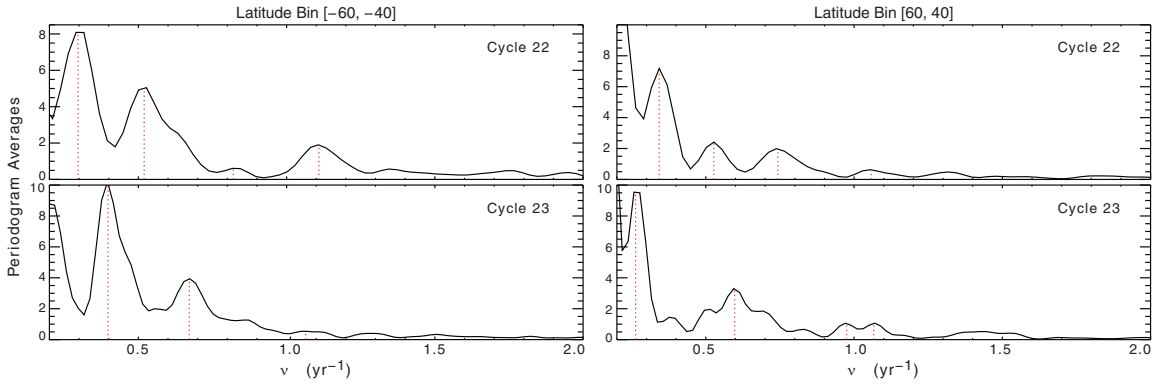}}
\caption{This figure shows a sample of four periodograms selected to be for the latitudes between 
40$^\circ$ and 60$^\circ$N/S.  The upper figures are for cycle 22 while the lower figures are for cycle 23.  The identified frequencies are shown by the vertical dashed lines (red in the on-line version).
}
\label{figureseven}
}
\end{center}
\end{figure}
\begin{figure}
\begin{center}
\parbox{6.0in}{
\resizebox{6.0in}{!}{\includegraphics{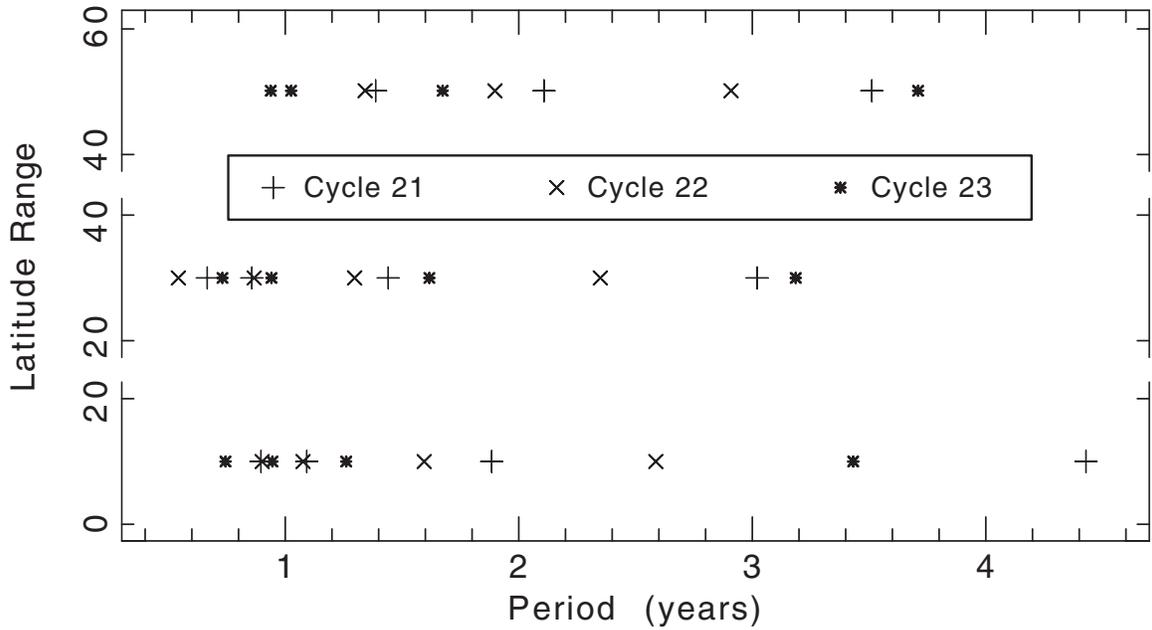}}
\caption{This figure shows the periods for all the peaks like those in figure~(\ref{figureseven}).  The
periods are sorted by the latitude band where they were found and differing plotting symbols are shown
for each cycle.
}
\label{figureeight}
}
\end{center}
\end{figure}

We searched for periodicities in the $dBr/dt$ map using the Lomb-Scargle periodiogram approach \citep{1982ApJ...263..835S}.  This method has the advantage that the time series does not need to be evenly sampled.  Consequently, we are able to study the time series without gap filling and without smoothing.  The resulting periodograms for single latitudes are erratic with the peak location being function of latitude
as well as a function of time when the original series is taken as several independent parts.  A set of sharp peaks that are not stable is often an indication that the lifetime of the underlying phenomenon is shorter than the full time series.  In order to better match the time series to the lifetime of the magnetic pattern, we sub-divided the series into three parts corresponding roughly with the times of cycles 21, 22 and 23.  In addition we have binned the latitudes into eight ranges, each of which is 20$^\circ$ wide.  This gives 24 separate periodograms.  There are few regularities among the periods found in these periodograms.  A sample of four is shown in figure~(\ref{figureseven}).  Each of the identified peaks has a vertical dashed line at the measured frequency.  All of the periods for identified peaks are shown as symbols on figure~(\ref{figureeight}) grouped according to latitude band.  The separate solar cycles are identified by the symbols according to the legend on the upper third of figure~(\ref{figureeight}).  There are no patterns among these periods and we conclude that we have not determined any organizing principle for the periodicity in the magnetic field fluctuations.

\section{Discussion}

\begin{figure}
\begin{center}
\parbox{5.5in}{
\resizebox{5.5in}{!}{\includegraphics{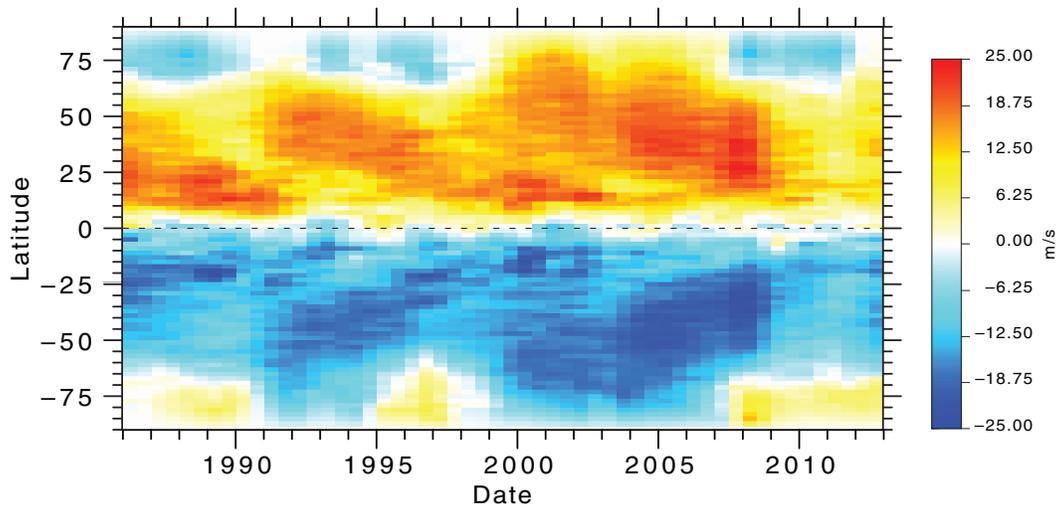}}
\caption{The map of the meridional circulation as a function of time and latitude derived with the method described by \citet{2010ApJ...725..658U}.  Each 6-month bin is the result of averaging over the first or second half of each year.  An even-odd feature present in the map was removed by using a $(v_i)_{\rm filtered}=0.25v_{i-1}+0.50v_i+ 0.25v_{i+1}$ temporal filter on each latitude.  The final time slot corresponding to the last 6 months of 2012 is in fact derived from only the months of July to Sept.
}
\label{figurenine}
}
\end{center}
\end{figure}

\chc
Maps of the sun's magnetic field like that of figure~(\ref{figurefour}) are one of the most powerful tools for the study
of the solar cycle and have been used regularly since the publication by \citet{1981SoPh...74..131H}.  Improved maps have been provided regularly with recent examples being found in \citet{2010LRSP....7....1H}, \citet{2012ApJ...749...27V} and \citet{2012SoPh..281..577P}.  In most of these prior maps the patterns known as surges are clearly visible and usually lead directly to
the reversal of the polar fields.  Also visible in these is what can be called a counter-surge a few years after the main surge during which the polarity of the previous cycle has a return to dominance in the mid-latitudes (30$^\circ$N/S to 60$^\circ$N/S) but with inadequate strength to overcome the newly dominant polar field.  These counter-surges are most evident in our figure~(\ref{figurefour}) and the map in \citet{2010LRSP....7....1H}.  

Although on the full magnetic map the counter-surges appear to be relatively minor fluctuations on the overall pattern of the solar cycle, they actually represent a fundamental problem.  According to the standard picture of the cycle involves the sequence: a) the polar field generates a toroidal field from differential rotation, b) the toroidal field strengthens until it can generate bipolar sunspot groups, c) the bipolar sunspot groups are oriented according to Joy's Law such that their trailing spots are poleward of the leading spot so that the trailing spot polarity becomes dominant in the mid-latitudes by virtue of the spot's favorable position relative to the pole, d) the trailing spot polarity reverses the previous dipole field so that the sequence can repeat.  As long as the Joy's Law tilt configuration of bipolar spots is stationary, the leading spot field cannot become dominant in the mid-latitudes independent of the number of spot groups present.  During a counter-surge the leading spot field is dominant indicating
that the spot tilt is opposite that given by Joy's Law. 
Although the counter-surges are rare and might be dismissed as anomalies, the ripples that are revealed by a close inspection of the magnetic field maps as enhanced in figures~(\ref{figuresixa}) and (\ref{figuresixb}) show that in fact the polarity dominance in the mid-latitudes is continually varying.  In the mid-latitudes where there are few spots, our average over a Carrington rotation gives us a quantitative measure of the degree to which the trailing polarity is in fact dominant.  Although the tilt angle is, for a long-term average, a function of the latitude \citep{2012ApJ...758..115L} the present study shows that the tilt angle is not steady. 
 
As is best seen in the time derivative plot of figure~(\ref{figuresixa}) the magnetic ripples alternate between reducing and enhancing the existing dipole structure with the cumulative effect causing the dipole component reversal.  We can understand how the changes in the dominant magnetic field in mid-latitudes are related to sunspot emergence by using the Babcock-Leighton model as extended by \citet{1991ApJ...383..431W}.  According to this model the time derivative
 $\partial B_s/\partial t$ has the form of a sum of an advection term, a diffusion term and a source term \citep[see equations 5 and 6]{1991ApJ...383..431W}.  The form of the source term, $S$, is:
\begin{eqnarray}
S(R_\sun,\theta,t) \propto \rat {1}{\sin\theta} \rat {\partial}{\partial \theta}\left(\rat {B_\phi a \sin\gamma}{\tau}\right)
\end{eqnarray}
where $B_s$ is the surface field strength which we identify with $B_r$, $B_\phi$ is the toroidal field strength in the sub-surface zone where sunspots appear, $a$ is the absolute value of 
pole separation between the components of a bipolar magnetic region, $\gamma$ is the Joy's Law tilt angle and $\tau$ is the time scale for the features to erupt.  The coefficient of proportionality is the ratio of several constant geometric factors.  The dominant polarity in the active zone is generated by the source term from sunspot groups and this polarity then spreads poleward as a result of the advection and diffusion terms.  A basic property of each sunspot cycle is the structure of the deep seated toroidal field $B_\phi$ which does not change sign until the next cycle.
Consequently, the alternating sign of $dB_r/dt$ can only be due to an alternating sign of $\gamma$.  Thus the relative orientation of the
spots as defined by the Joy's Law tilt angle $\gamma$ must be opposite from normal during periods of reversed sign of $dB_r/dt$.  

The wave pattern of the magnetic ripples drifts in latitude as if their wave pattern is being advected by the
meridional circulation.  The actual ripple pattern is complex in many places and simple in a few.  In those
simple places the peaks and troughs can be tracked so that it is possible to determine a pattern drift 
velocity.  While that pattern drift velocity is comparable in magnitude to the surface Doppler meridional 
circulation velocity reported by \citet{2010ApJ...725..658U}, the match in detail is quite imperfect.  For reference a 2-d map of
the meridional circulation velocity found by \citet{2010ApJ...725..658U} is shown in figure \ref{figurenine}.  The dual cell structure
on this map represents a particular problem since it implies a zero drift rate along the boundary between the dominant poleward circulation cell
and the polar cell where the circulation is equatorward. There are no cases where the drift of the magnetic ripple stagnates at this boundary.  This disagreement indicates that the pattern
drift is not caused by surface mass motions alone and is a consequence of a more complex interaction such as more deeply seated mass motion, magnetic forces or flux emergence.  An alternate determination of the meridional circulation in a recent study by \citet{2012ApJ...761L..14R} based on pattern tracking of small-scale magnetic features found that the reverse cell at high latitudes was absent during the period from Apr.\ 2010 to Feb.\ 2011.\chcend\

A striking property of the ripples is their long duration -- lifetimes of two to three years and extending over latitudes starting at 20$^\circ$ and ending often at the pole.  Some of the ripples start at the
equator and extend into both hemispheres although this behavior is less common. 
The general time scale of the ripples is similar to that of the Quasi-Biennial Oscillations that
have been discussed by a variety of authors \citep{2000Sci...287.2456H,2003ESASP.535....3B,2005SoPh..227..155K,2009SoPh..257...61J,2012MNRAS.420.1405B}.  The cause of these QBO's is not understood but the presence of a periodicity in this range in rotation rates, especially deep in the convection zone, suggests that the QBO's are related to deep-seated processes governing the solar cycle.

It is interesting that in the northern hemisphere there are three wider than average bands of $dB_r/dt$ of -, + and - sign starting at about 60$^\circ$N at times of 1978, 1988 and 1998 and ending near the poles at times of 1980, 1990 and 2000.  These bands are of long enough temporal duration that they do not show up in the $\delta B_r$ plot of figure \ref{figuresixb} but rather are seen in the smooth $B_r$ part of the figure and immediately preceed the reversal of the polar field.  These correspond roughly to the times when coronal features undergo a rapid migration from mid-latitudes to the poles called the ``rush to the poles'' \citep{2011SoPh..274..251A}.  The stronger-than-average ripples in the rate of magnetic field change are evidently associated with these coronal processes while the majority of the ripples do not have an impact on this global structure.  In addition, it is noteworthy that the rotation rate of the coronal features also discussed by \citet{2003SoPh..213...23A} is significantly different from the rate derived from the Doppler shift in the polar regions.  This is consistent with the idea that the magnetic field motions and evolution is not simply a consequence of advection by the matter on the solar surface.

\section{Conclusions}
We have presented improvements to the derivation of solar surface magnetic field strengths as measured at the Mt.\ Wilson Observatory.  These improvements include the deduction of a tilt angle $\zeta$ for the orientation of the field lines near the poles and the filling of unseen portions of the poles using a method based on the techniques of \citet{2009PhDT.........1T}.  These improvements substantially reduce the annual variation in the line-of-sight field due to the changing viewing angle (the $b_0$ effect).  Owing to the low-noise characteristics of the final magnetic field map, we are able to identify the presence of ubiquitous ripples in the magnetic field that generally start at low latitudes and propagate to the poles.  The motion of these ripples cannot be due to advection at the surface alone due to the apparent inverse relationship between the surface meridional circulation and the apparent travel speed of the features.  It is likely that in addition to the hydrodynamic influence of the convection zone matter, large-scale 
magnetic forces also play a role in the structure and motion of the ripples.

The ripples in the $dB_r/dt$ map clearly are associated with the
process that reverses the dipole during the solar cycle.   
Because we are considering averages
over full circles at constant latitude, the $B_r$ value we find represents 
an axially symmetric pattern.  \ch At the time of the
dipole reversal the value of $dB_r/dt$ is larger than average (note that the $\delta B_r$ part of figure \ref{figuresixb} does not show this effect because its duration is longer that 2.5 years and the deviation shows up on the $B_r$ smoothed panel instead) \chend and is associated
with the ``rush-to-the-poles'' structural changes in the corona.  A part of dynamo models of the solar cycle is the $\alpha$-effect [see for example the review by \citet{2010LRSP....7....3C}] whereby rising portions
of the underlying toroidal magnetic field are twisted in such a way that the
global dipole field is reversed.  This twisting is manifested as the Joy's Law
tilt of active regions \chb and the magnetic ripple responds to changes in this Joy's Law tilt.  \chbnd

It is possible that there are \ch relationships between the magnetic ripples and other solar cycle phenomena such as flare production \citep{1984Natur.312..623R}, sunspots \citep{2002A&A...394..701K} or geophysical indices \citep{2005SoPh..227..155K}.  Since the \chb ripple \chbnd structure is large scale and long lasting, it could be associated with structures in the heliospheric magnetic field that are manifested in the geophysical indices.  However, in the absence of a specific physical
model as a basis for a search for such relationships, temporal concident and lagged cross-correlation studies may
provide the best available tool.  Because the magnetic ripples do not show any clear and stable periodicities, a coincidence of periods is only a rough guide to a search that would need 
confirmation through a cross-correlation approach.\chend

\acknowledgments
We thank the referees for helpful comments that have improved this paper.
We thank John Boyden for his management of the data acquisition software and for maintaining a variety of the hardware and electronic systems at the 150-foot tower telescope.  We also thank Steve Padilla and Micheal Tu for their continued service in acquiring the observations.  We also thank the many additional observers who have obtained the data over the years, especially Larry Webster and Pamela Gilman.  We thank Todd Hoeksema, Peter Gilman and Jack Harvey for reading an early draft of this paper and providing helpful suggestions.  This work has been supported recently by the NSF through grant AGS-0958779 and NASA through grants NNX09AB12G and HMI subcontract 16165880.  Over the years additional funding has come from these two agencies as well as the ONR and NOAA.

\noindent e-mail: \email{ulrich@astro.ucla.edu}.

\end{document}